\documentclass[aps, prb, reprint, superscriptaddress, floatfix]{revtex4-2}

\pdfoutput=1

\usepackage{xcolor}
\usepackage{booktabs}
\usepackage{graphicx}
\usepackage{dcolumn}
\usepackage{amssymb}
\usepackage{amsmath}
\usepackage{hyperref}
\hypersetup{colorlinks=true, citecolor=blue, linkcolor=blue, urlcolor=blue}
\usepackage[Symbolsmallscale]{upgreek}
\usepackage{enumitem}

\newcommand{\emc}{e$\upmu$c}
\newcommand{\br}{\textbf{r}}

\newcommand{\erho}{\rho_\mathrm{e}}

\newcommand{\mrho}{\rho_\upmu}
\newcolumntype{d}[1]{D{.}{.}{#1}}
\newcommand{\cmmnt}[1]{\ignorespaces} 

\begin{document}

\title{Comment on “Two-component density functional theory study of quantized muons in solids”}

\author{Mohammad Goli}
\email{m{\_}goli@ipm.ir}
\affiliation{School of Nano Science, Institute for Research in Fundamental Sciences (IPM), Tehran, 19395-5531, Iran}

\author{Nahid Sadat Riyahi}
\affiliation{Department of Physical and Computational Chemistry, Shahid Beheshti University, Evin, Tehran, 19839-69411, Iran}

\author{Shant Shahbazian}
\email{sh{\_}shahbazian@sbu.ac.ir}
\affiliation{Department of Physics, Shahid Beheshti University, Evin, Tehran, Iran}

\date{\today}

\begin{abstract}
In [Phys. Rev. B 107, 094433 (2023)], Deng et al. have proposed an electron-muon correlation functional within the context of the two-component density functional theory (TC-DFT) for crystals/molecules containing positively charged muons. In order to verify its performance, we applied the functional in conjunction with the B3LYP, as a hybrid electronic exchange-correlation functional, to a benchmark set of molecules. The results demonstrate that the proposed functional is not capable of reproducing the correct one-muon densities as well as some other key properties like muon’s kinetic energy, the total energies and the mean muonic bond lengths. Using the muonium atom in a double-harmonic trap as a model we also demonstrate that the successful reproduction of the electron-muon contact hyperfine coupling constants by Deng et al. is probably the result of error cancellations. We also discuss some theoretical intricacies with the very definition of the electron-muon correlation energy within the context of the TC-DFT that must be taken into account in future efforts to design electron-muon correlation functionals. 
\end{abstract}


\maketitle

In a recent paper \cite{deng_twocomponent_2023}, Deng et al. have applied the two-component density functional theory (TC-DFT) to some crystalline structures where a positively charged muon, hereafter called just muon, is implanted within the unit cell. To implement the TC Kohn-Sham (TC-KS) equations computationally, they have proposed a new electron-muon correlation functional (EMCF), see Eq. (15) in Ref. \cite{deng_twocomponent_2023}, hereafter called DYPZPY, by reparameterizing the \emc-1 functional proposed by our group some time ago \cite{goli_twocomponent_2022}. The electron-muon correlation energies (EMCEs) derived from the quantum Monte Carlo calculations of the TC-EM homogeneous gas system (TC-EM-HGS) were employed for the reparameterization. A novel zero-distance enhancement factor (ZDEF) was also derived to compute the electron-muon contact hyperfine coupling constants (EM-CHCCs). While the study conducted by Deng et al. is a step forward in applying the TC-DFT to muonic species at condensed phases, we believe it suffers from some theoretical and computational ambiguities that must be clarified. In this comment, we will try to dig into the details of these ambiguities which may hopefully pave the way for further developments in the TC-DFT realization of muonic species (see Sec. S1 A of the Supplemental Material (SM) for further discussion on the conceptual background of the muonic TC-DFT \cite{supp_mat}). According to our previous study \cite{goli_twocomponent_2022}, we proposed the following goals for a successful application of the muonic TC-DFT:
\begin{enumerate}
\item To be capable of reproducing the one-muon density and muonic kinetic energy in an accurate, quantitative manner.
\item To be capable of reproducing the potential energy surface of muonic systems, yielding the corresponding optimized geometries as well as the vibrational spectrum of clamped nuclei without referring to any prior calculations within the single- or double-adiabatic approximation framework.
\end{enumerate}
According to the study of Deng et al., one may add a third goal namely:
\begin{enumerate}[resume]
\item To be capable of reproducing the EM-CHCCs in an accurate, quantitative manner.
\end{enumerate}
If the third goal materialized, it would be feasible to computationally reproduce the results of various muon spin spectroscopies \cite{nagamine_introductory_2003}. These spectroscopies are currently the main experimental window into the characterization of the muonic molecular systems and those muonic species formed in condensed phases \cite{heffner_muon_1984,roduner_muon_1997,blundell_spinpolarized_1999,blundell_muonspin_2004,mckenzie_positive_2013,nuccio_muon_2014,hillier_muon_2022}. As will be discussed, to achieve this goal one needs to have/approximate the exact ZDEF that may go beyond TC-DFT. Let us stress that at the heart of all three goals is the correct reproduction of the one-muon density (vide infra) since both the zero-point energies and EM-CHCCs are intrinsically linked to this density.

\section{Electron-muon correlation energy}
The concept of electron-electron correlation in purely, i.e. single-component (SC), electronic systems is a complex but well-studied subject both within the context of wavefunction-based and DFT methods \cite{baerends_quantum_1997,helgaker_molecular_2000,fulde_correlated_2012}. In contrast, for a general TC quantum system composed of electrons and a class of positively charged particles (PCP), e.g. muons or positrons, there are three types of correlations: the electron-electron, the electron-PCP and the PCP-PCP correlations. In general, no universal recipe has yet been proposed to disentangle the energetic contribution of these three types of correlations, and based on the details of the method of computation, various separations are feasible \cite{swalina_alternative_2005,chakraborty_development_2008,udagawa_electronnucleus_2014,cassam-chenai_decoupling_2015,brorsen_nuclearelectronic_2015,cassam-chenai_quantum_2017,brorsen_alternative_2018,fajen_separation_2020}. Even if we constrain ourselves to the TC-DFT framework, more than a single way of separation of these correlations is feasible (see Sec. S1 B of the SM for the relevant theory and two unequal EMCE definitions \cite{supp_mat}).

Interestingly, within the special subfield of the TC-DFT of the positronic molecules and condensed phases, the possibility of alternative formulations of TC-DFT was recognized long ago \cite{chakraborty_electron_1983,nieminen_twocomponent_1985}, and the same argument is also applicable to the TC-DFT of the muonic systems. Deng et al. have not discussed their preferred definition of the EMCE in Ref. \cite{deng_twocomponent_2023}; however, in Eq. (14) of the paper, they have introduced what they have called the “two-body correlation energy” of the TC-EM-HGS. This is done by subtracting the sum of the total energies of the two SC-HGSs composed exclusively from electrons or muons from the total energy of the TC system: $\Delta E = E_\mathrm{total}(N_\mathrm{e},N_\mathrm{\upmu})-(E_\mathrm{total}(N_\mathrm{e})+E_\mathrm{total}(N_\mathrm{\upmu}))$. Assuming $ E_{\mathrm{e}\upmu\mathrm{c}} = \Delta E $, the EMCF is then parameterized by fitting the functional to the computed $\Delta E$ values. While at first glance this may seem a reasonable choice, upon further investigation it can be shown that this definition is not inevitably compatible with electron-muon correlation energies introduced within the framework of TC-DFT (see Sec. S1 B of the SM \cite{supp_mat}). 

Let us consider the case of Eq. (S4) in the SM \cite{supp_mat} as an illustrative example regarding the partitioning of the total exchange-correlation energy. Applying the charge neutrality condition to an SC-HGS yields the total energy within the context of the KS-DFT: $E_{\mathrm{total}}[n_\upalpha] = T^{s}[n_\upalpha]+E_{\mathrm{\upalpha xc}}[n_\upalpha]$, wherein $n_\upalpha (\br_\upalpha) = N_\upalpha \sum_{\mathrm{spin}} \int d \br_{2} \cdots \int d \br_{N_\upalpha} |\psi_{\upalpha}|^2 $ is the usual one-electron/muon density, $\upalpha=\mathrm{e}/\upmu$ (for details see Ref. \cite{loos_uniform_2016}). It is also straightforward to derive the total energy of the TC-HGS by applying the charge neutrality condition: $ E_{\mathrm{total}} \left [\erho,\mrho \right ] = T_\mathrm{e}^{s}\left [ \erho \right ]+T_\upmu^{s}\left [ \mrho \right ]+ E_\mathrm{xc}^{\mathrm{total}}\left [ \erho,\mrho \right ]$ where $\rho_{\upalpha} (\br_\upalpha) = N_\upalpha \sum_{\mathrm{spin}} \int d \br_{2,\upalpha} \cdots \int d \br_{N_\upalpha,\upalpha} \int d \br_{1,\upbeta} \cdots \int d \br_{N_\upbeta,\upbeta} | \psi_{\upalpha \upbeta} |^2 $ and $\upbeta=\mathrm{e}/\upmu$. With these results at hand $\Delta E$ is:   
\begin{align}\label{eq:1}
\Delta E = & \left( T_\mathrm{e}^{s}\left [ \erho \right ] -T_\mathrm{e}^{s}\left [ n_\mathrm{e} \right ] \right) + \left( T_\upmu^{s}\left [ \mrho \right ] -T_\upmu^{s}\left [ n_\upmu \right ] \right) \nonumber \\
& + \left( E_\mathrm{exc}\left [ \erho,\mrho \right ] - E_\mathrm{exc}\left [ n_\mathrm{e} \right ] \right) \nonumber \\
& + \left( E_\mathrm{\upmu xc}\left [ \erho,\mrho \right ] - E_\mathrm{\upmu xc}\left [ n_\upmu \right ] \right) + E_\mathrm{e \upmu c}\left [ \erho,\mrho \right ].
\end{align}
In general, there is no theoretical basis to believe that the contributions of four parentheses on the r.h.s. of Eq. (\ref{eq:1}) are negligible or somehow cancel each other when $\rho_\upalpha=n_\upalpha$ since we are faced with strongly interacting systems, and even it is hard to assess the relative importance of all five terms from mere theoretical arguments. However, it is probably reasonable to claim for cases where $N_\mathrm{e} \gg N_\upmu$ or $N_\upmu \gg N_\mathrm{e}$, similar to the case of a positron in an electron gas system \cite{drummond_quantum_2011}: $\Delta E \approx E_\mathrm{e \upmu c}\left [ \erho,\mrho \right ]$, but it is hard to justify this result for cases in which $N_\mathrm{e} \sim N_\upmu$. Taking the fact that Deng et al. have used numerical data from the latter region (see the description below Eq. (14) in Ref. \cite{deng_twocomponent_2023}) to parameterize the DYPZPY functional, acquiring contamination from the first four terms on the r.h.s. of Eq. (\ref{eq:1}) seems inevitable. As stressed, estimating the overall extent of this contamination is difficult, and without a detailed study on the relative importance of these terms, the reliability of the DYPZPY functional to reproduce the “pure” electron-muon correlation energies of the TC-EM-HGS remains unclear.      

\begin{figure}[t]
\includegraphics[width=\columnwidth]{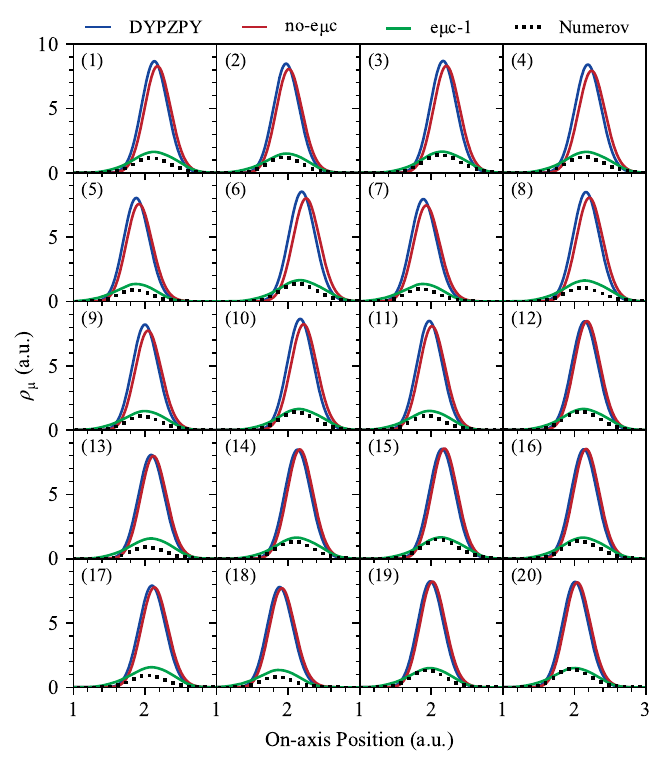}%
\caption{On-axis depictions of the one-muon density computed at the B3LYP/pc-2//DYPZPY/14s14p14d, B3LYP/pc-2//no-{\emc}/14s14p14d, B3LYP/pc-2//{\emc}-1/14s14p14d and Numerov levels of theory (for computational details see Ref. \cite{goli_twocomponent_2022}). The on-axis position is the direction that is parallel to the vector linking the muon attachment site to the maximum of $\mrho$ obtained at the corresponding level of theory. The muon attachment site is placed at the center of the coordinate system. The numbers given in parentheses are the numbering scheme used to label each molecule (for details of molecular structures see Figure 1 in Ref. \cite{goli_twocomponent_2022}).}
\label{fig:1}
\end{figure}

\section{Performance of the DYPZPY correlation functional}
Let us for the moment dismiss the contamination problem with the DYPZPY functional and try to evaluate its performance. One may be critical to whether the TC-EM-HGS is a proper model for typical inhomogeneous muonic systems in which the muon is a well-localized particle (vide infra). Nevertheless, as also rightfully assumed by Deng et al. \cite{deng_twocomponent_2023}, nothing in principle forbids one to try various strategies to propose an EMCF. Thus, in this section, we try to evaluate the DYPZPY performance computationally by applying it to the molecular set disclosed in our previous study \cite{goli_twocomponent_2022}. Let us just stress that our proposed set contains twenty muonic organic molecules including both open- and closed-shell species. A detailed vibrational analysis demonstrated that in this set the muon’s vibrational modes to a large extent are decoupled from those of the nuclei; thus, treating the muon and electrons on equal footing within the context of the TC-DFT is justified. For a comparative study, we used the DYPZPY functional for both electrons and muon (discarding the modified electronic correlation potential in Eq. (17) of Ref. \cite{deng_twocomponent_2023}), and followed the same analysis that was done for evaluating the \emc-1 performance in our previous study \cite{goli_twocomponent_2022}. The results are summarized in Fig. \ref{fig:1} and Table \ref{tab:1}.

Figure \ref{fig:1} depicts $\mrho$ computed along the muon's bonding axis. The TC-DFT results using the DYPZPY and \emc-1 functionals and without any EMCF, called no-{\emc} hereafter, are offered in this figure. For comparison, $\mrho$ was also computed using the generalized matrix Numerov method at the double-adiabatic level \cite{goli_twocomponent_2022} as the reference/exact data. The B3LYP hybrid exchange-correlation functional was used for electrons in these computations. Even a glance reveals the fact that in contrast to \emc-1 which almost reproduces the exact results derived by the Numerov method, the results obtained using DYPZPY are virtually indiscernible from those derived using no-{\emc}.

Clearly, DYPZPY is not able to remedy the “overlocalization” of $\mrho$ produced at the no-{\emc} level, which is the first goal of the TC-DFT articulated previously. Let us stress that similarly in the case of proton the main motivation of designing the epc series of the electron-proton correlation functionals, i.e. epc-17, epc-18 and epc-19, was to overcome the same overlocalization dilemma \cite{brorsen_alternative_2018,yang_development_2017,brorsen_multicomponent_2017,tao_multicomponent_2019}. To have a more detailed picture of the performance of DYPZPY, the mean error (ME) and root mean square error (RMSE) of the total energies, the muonic kinetic energies and the mean muonic bond lengths with respect to the Numerov results, as well as the root mean square deviation (RMSD) of the computed $\mrho$ from the densities obtained using the Numerov method, are offered in Table \ref{tab:1} and compared with those derived at the \emc-1 and no-{\emc} levels of theory. The individual values of the computed properties and the optimized geometries of the molecular structures are also gathered in Tables S1 and S2 of the SM, respectively \cite{supp_mat}. Once again, the ME and the RMSE computed using DYPZPY are much more similar to those computed at the no-{\emc} level and far from those derived using \emc-1. We conclude that the DYPZPY is incapable of reproducing the main observable of interest in muonic molecules; thus, the reparameterization of \emc-1 does not seem to yield a better match to the reference data. 

\begin{table}[t]
  \centering
  \caption{ME and RMSE of the total energies $E$ ($E_h$), muonic kinetic energies $K_{\upmu}$ ($E_h$), muonic bond length expectation values $\langle r_{\upmu} \rangle$ (\r{A}), and RMSDs of $\mrho$ (a.u.) computed at the TC-DFT levels.}
  \begin{ruledtabular}
    \begin{tabular}{l d{2.3}d{1.3}d{1.3}}
      &  \multicolumn{1}{c}{e$\upmu$c-1}\footnotemark[1]    & \multicolumn{1}{c}{DYPZPY}\footnotemark[2]  & \multicolumn{1}{c}{no-e$\upmu$c}\footnotemark[3]   \\ \hline
      &   \multicolumn{3}{c}{ME}   \\ \cmidrule{2-4}
      $E$                             & -0.005 & 0.059 & 0.077 \\
      $K_{\upmu}$                     & 0.002 & 0.030 & 0.028 \\
      $\langle r_{\upmu} \rangle$     & 0.005 & 0.048 & 0.070 \\
      RMSD                            & 0.030 & 0.153 & 0.151 \\
      &  \multicolumn{3}{c}{RMSE}  \\  \cmidrule{2-4}
      $E$                             & 0.008 & 0.059 & 0.077 \\
      $K_{\upmu}$                     & 0.003 & 0.030 & 0.028 \\
      $\langle r_{\upmu} \rangle$     & 0.014 & 0.052 & 0.073 \\
    \end{tabular}%
    \end{ruledtabular}
    \footnotetext[1]{Obtained at the B3LYP/pc-2//{\emc}-1/14s14p14d level of theory from Ref. \cite{goli_twocomponent_2022}.}
    \footnotetext[2]{Computed at the B3LYP/pc-2//DYPZPY/14s14p14d level of theory.}
    \footnotetext[3]{Obtained at the B3LYP/pc-2//no-{\emc}/14s14p14d level of theory from Ref. \cite{goli_twocomponent_2022}.}
  \label{tab:1}%
\end{table}

\begin{figure}[t]
\includegraphics[width=\columnwidth]{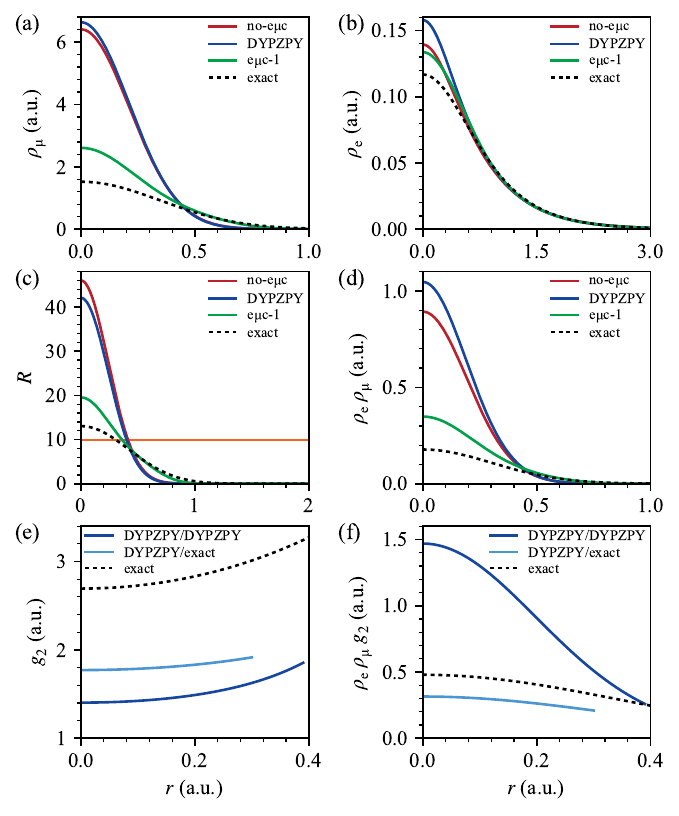}%
\caption{Exact and TC-DFT computed results for the muonium atom in the double-harmonic trap model. (a) $\mrho$, (b) $\erho$, (c) the $R$ ratio, (d) the on-top density product, $\erho \mrho$, (e) $g_2$ computed from the exact $\mrho$, indicated by DYPZPY/exact, and from $\mrho$ derived using the DYPZPY functional, indicated by DYPZPY/ DYPZPY, and the exact ZDEF, (f) the on-top product $\erho \mrho g_2$.}
\label{fig:2}
\end{figure}

Interestingly, Deng et al. did not include the correlation potential of DYPZPY in the muonic KS equation but it has been incorporated only for the electronic KS equation; the same strategy also has been used in a more recent study (see Eq. (2) in Ref. \cite{pan_fm_2023}). Nevertheless, even in this case, the DYPZPY-derived electronic correlation potential has been modified heavily in cases where the $\mrho$ dominates $\erho$ (see Eqs. (16) and (17) in Ref. \cite{deng_twocomponent_2023}). Such \textit{ad hoc} modifications of the correlation potential are not only computationally troublesome without having access to the exact $ E_\mathrm{xc}^{\mathrm{total}}$ value but also point to the fact that Deng et al. did not find the DYPZPY derived original electronic correlation potential proper enough to be used directly in their subsequent computational studies. This incomplete semi-empirical KS-DFT approach would lead to an inconsistent KS-DFT procedure rendering the whole approach arbitrary to a large extent.

\section{Performance of the zero-distance enhancement factor}
As emphasized, the main target of Deng et al. was reproducing the experimentally derived EM-CHCCs in crystalline structures, and based on the results of Table V in Ref. \cite{deng_twocomponent_2023}, they seem to be successful in this goal. This is incompatible with the overlocalized nature of their computed $\mrho$ as detailed in the previous section and also observable in Fig. 7 of Ref. \cite{deng_twocomponent_2023} where the heights of the maxima of $\mrho$ in various crystalline structures are $\sim$9-11 in atomic units. These values are almost in the same range of the heights of the maxima of $\mrho$ depicted in Fig. \ref{fig:1}, $\sim$8-9 in atomic units, at no-{\emc} level and TC-DFT calculations using the DYPZPY functional but far from the Numerov reference data which are well-reproduced using the \emc-1 functional, $\sim$1-2 in atomic units. Evidently, the reason behind the success of Deng et al. is not the well-reproduction of the exact $\mrho$; thus, we will try to delve into the intricacies of this problem in the rest of this section.   

The following integral was used by Deng et al. to compute the EM-CHCCs: $A=C \int d \br \erho^{s}(\br) \mrho (\br) g(0;\br) $, in which $\erho^{s}(\br) = \erho^{\upalpha}(\br) - \erho^{\upbeta}(\br)$ is the electronic spin-density and $C$ contains all the constants (for details see Eqs. (8) and (9) in Ref. \cite{deng_twocomponent_2023}). Also, $g(0;\br)$ is the ZDEF which is: $g(0;\br)=\frac{\rho^{(2)}(\br_\mathrm{e},\br_\mathrm{\upmu})}{\erho(\br_\mathrm{e}) \mrho(\br_\mathrm{\upmu})}\big|_{\br=\br_\mathrm{e}=\br_\mathrm{\upmu}}$, where $\rho^{(2)}(\br_\mathrm{e},\br_\mathrm{\upmu}) = N_\mathrm{e} N_\mathrm{\upmu} \sum_{\mathrm{spin}} \int d \br_{2,\mathrm{e}} \cdots \int d \br_{N_\mathrm{e},\mathrm{e}} \int d \br_{2,\mathrm{\upmu}} \cdots \int d \br_{N_\mathrm{\upmu},\mathrm{\upmu}} | \Psi_{\mathrm{e\upmu}} |^2$ is the EM pair density. Deng et al. used the data from the TC-EM-HGS computed as various $\mrho$ and $\erho$ values to fit their density-dependent ZDEF, $g(0;\erho,\mrho)$. The resulting ZDEF, called DYPZPY-ZDEF hereafter, is a complicated functional of $\mrho$ and $\erho$ and is given by three separate rules, called $g_1$, $g_2$ and $g_3$, each employed in a different ratio of $R(\br)= \mrho(\br)/\erho(\br)$ (for details see Eqs. (11), (12) and (13) in Ref. \cite{deng_twocomponent_2023}). The explicit form of $g_1$ has not been given, and $g_3$ is the rule if $R<0.2$ which applies to the regions of space far from the maxima of $ \erho $ and $\mrho $ probably having a marginal effect on the value of EM-CHCC integral. Therefore, herein we only concentrate on $g_2$ which is the rule if $R>10$ and only depends on $\mrho$. Assuming $\erho^{s}(\br)$ has been rather accurately computed by Deng et al., the only way to compensate for the overlocalization of $\mrho (\br)$ and to deduce the correct value for the EM-CHCCs is some kind of error cancellation through $g(0;\erho,\mrho)$. To verify this hypothesis, similar to the approach adopted by Deng et al. in Ref. \cite{deng_twocomponent_2023}, we employ the muonium atom as a model.

Deng et al. used the muonium atom, composed of an electron and a muon, in a vacuum as a model to tune the parameter of their proposed semi-empirical electronic correlation potential Eq. (17) to reproduce the experimental EM-CHCC \cite{deng_twocomponent_2023}. To verify the performance of $g_2$, we also employ the muonium atom but instead of a vacuum, we placed it in a double-harmonic trap model to eliminate the obstacles that arise in dealing with the center of mass translational motion \cite{bochevarov_electron_2004}. Accordingly, one may derive the exact $\mrho$ and $\erho$, corresponding $g_2$ and the exact ZDEF, all depicted in Fig. \ref{fig:2}; the theory and details of the TC-DFT computations for the double-harmonic trap model are provided in Sec. S1 C of the SM \cite{supp_mat}.

It is evident from panel (a) of the figure that TC-DFT derived $\mrho$ employing the DYPZPY functional, in contrast to the one derived employing \emc-1, is quite disparate from the exact $\mrho$ and is virtually superimposable on $\mrho$ derived at the no-{\emc} level. While in the case of $\erho$, depicted in panel (b), the disparity between the computed densities is much smaller compared to $\mrho$, and strangely, $\erho$ derived using the DYPZPY functional seems to be the least accurate one. The $R$ ratio given in panel (c) demonstrates that $g_2$ is the rule of DYPZPY-ZDEF for $r<0.4$ Bohr. Let us now consider the local behavior of the integrand used to compute the EM-CHCC. Neglecting the ZDEF, the remaining part of the integrand is the on-top product of the densities, $\erho\mrho$, and as depicted in panel (d), it is unusually overlocalized, even more than that derived at the no-{\emc} level when the DYPZPY functional is used in TC-DFT calculations. Panel (e) reveals that $g_2$ is always substantially smaller than the exact ZDEF at all the considered distances if $\mrho$ is derived employing the DYPZPY functional, and indeed compensating to some extent for the overlocalized product of densities. Interestingly, as is seen in panel (f), using the exact $\mrho$ instead of that derived from TC-DFT calculation with the DYPZPY functional in the total integrand $\erho \mrho g_2$ yields an even better match with the exact ZDEF demonstrating that $g_2$ is not intrinsically flawed. The radial distribution functions of the densities and their products are also given in Fig. S1 of the SM showing a general contraction trend as well as overlocalization in the DYPZPY functional results compared to the \emc-1 and exact data \cite{supp_mat}.

\section{Conclusion}
To achieve the triple goals of muonic TC-DFT, the correct reproduction of $\mrho$ and a properly designed ZDEF are both inevitable. None seem to be easy targets since our knowledge of the electron-muon correlation effects is yet in its infancy. Accordingly, trying to find model systems to design new EMCFs could be a proper strategy, as employed by Deng et al., but the performance of the emerging functionals must be evaluated with great care. We believe that none of the EMCFs has reached the triple goals concomitantly yet; the race to find the holy grail of the muonic TC-DFT is still ongoing.

\begin{acknowledgments}
S.S. would like to acknowledge generous access to the computational resources of the SARMAD Cluster at Shahid Beheshti University.
\end{acknowledgments}


%

\end{document}


\title{Supplemental Material: Comment on “Two-component density functional theory study of quantized muons in solids”}

\author{Mohammad Goli}
\email{m{\_}goli@ipm.ir}
\affiliation{School of Nano Science, Institute for Research in Fundamental Sciences (IPM), Tehran, 19395-5531, Iran}

\author{Nahid Sadat Riyahi}
\affiliation{Department of Physical and Computational Chemistry, Shahid Beheshti University, Evin, Tehran, 19839-69411, Iran}

\author{Shant Shahbazian}
\email{sh{\_}shahbazian@sbu.ac.ir}
\affiliation{Department of Physics, Shahid Beheshti University, Evin, Tehran, Iran}

\date{\today}

\maketitle

\onecolumngrid

\tableofcontents
\listoftables
\listoffigures
\clearpage

\section{Theoretical and computational details}
\subsection{Conceptual background of the muonic TC-DFT }
The idea of applying DFT to the muonic species, crystals or molecules, is not new and has been reviewed recently elsewhere \cite{bonfa_computational_2016,blundell_dft_2023}. However, in the usual studies based on the adiabatic approximation muon’s motion is assumed to be decoupled from that of electrons’ and it is treated as a light nucleus. In such a framework muon’s motion may be assumed to be coupled with other nuclei, called single-adiabatic approximation, or, it can be treated as a single particle within an effective field produced by both electrons and nuclei, called double-adiabatic approximation \cite{bonfa_efficient_2015}. In contrast, the basic premise behind the recently formulated TC-DFT is to incorporate the electrons’ and muon’s dynamics in a single Schr\"{o}dinger equation decoupled from the heavy nuclei \cite{goli_twocomponent_2022}. Such formulation makes it possible, at least in principle, to incorporate the non-adiabatic couplings between electrons and the muon into DFT equations. Whether such non-adiabatic effects may have a significant effect on the observables which are of interest in the field of muonic physics and chemistry remains to be seen. Nevertheless, in the meantime, one expects that the TC-DFT must at least be able to recover the results derived from DFT calculations done within the adiabatic or double-adiabatic approximations. 

\subsection{Electron-muon correlation energy theory}
Let us assume a general TC interacting system composed of $N_\mathrm{e}$/$N_\mathrm{p}$ number of electrons/PCPs within the electric field generated by $M$ number of clamped nuclei, described by the following TC Hamiltonian:
\begin{equation}
\hat{H}=\hat{T}_\mathrm{e}+\hat{T}_\mathrm{p}+\hat{V}_\mathrm{ee}+\hat{V}_\mathrm{ep}+\hat{V}_\mathrm{pp}+\hat{V}_\mathrm{e}^\mathrm{ext}+\hat{V}_\mathrm{p}^\mathrm{ext}.
\end{equation}
Each term is detailed as follows in atomic units: $\hat{T}_\upalpha=-\frac{1}{2 m_\upalpha}\sum_{i=1}^{N_\upalpha}\nabla_{\upalpha,i}^2$, $\hat{V}_{\upalpha \upbeta}=\sum_{i=1}^{N_\upalpha}\sum_{j=1}^{N_\upbeta}\frac{q_{\upalpha}q_{\upbeta}}{|\br_{\upalpha,i}-\br_{\upbeta,j}|}$, $\hat{V}_{\upalpha}^{\mathrm{ext}}=\sum_{i=1}^{N_\upalpha} \nu_{\upalpha}^{\mathrm{ext}}(\br_{\upalpha,i})$, and $\nu_{\upalpha}^{\mathrm{ext}}(\br_{\upalpha,i})=q_{\upalpha} \sum_{k=1}^{M}\frac{Z_{k}}{|\br_{\upalpha,i}-\bR_{k}|}$, wherein $\upalpha$ and $\upbeta$ are indices for e or p, and, $m_\upalpha$ and $q_{\upalpha}$ are the mass and charge of the electron/PCP, respectively, while $Z_{k}$ and $\bR_{k}$ are the charge and position vector of the $k$th clamped nucleus, respectively. It is straightforward to demonstrate that the TC Hohenberg-Kohn theorem and the generalized constraint Levy search implies (for details see subsection 9.6 in Ref. \cite{parr_densityfunctional_1994}):
\begin{align}\label{eq:2}
E_{\mathrm{ground}} = &\underset{\erho,\prho}{\mathrm{Min}} \; E \left [ \erho,\prho \right ], \nonumber \\
E \left [ \erho,\prho \right ] = & F \left [ \erho,\prho \right ] + \int d\br_{\mathrm{e}} \nu_{\mathrm{e}}^{\mathrm{ext}}(\br_\mathrm{e}) \erho (\br_\mathrm{e}) + \int d\br_{\mathrm{p}} \nu_{\mathrm{p}}^{\mathrm{ext}}(\br_\mathrm{p}) \prho (\br_\mathrm{p}), \nonumber \\
F \left [\erho,\prho \right ] = & \underset{ \Psi_{\mathrm{ep}} \rightarrow \erho,\prho}{\mathrm{Min}} \left \langle \Psi_{\mathrm{ep}} \left | \hat{T}_\mathrm{e}+\hat{T}_\mathrm{p}+\hat{V}_\mathrm{ee}+\hat{V}_\mathrm{ep}+\hat{V}_\mathrm{pp} \right | \Psi_{\mathrm{ep}} \right \rangle \nonumber \\
= &T_\mathrm{e}\left [ \erho,\prho \right ]+T_\mathrm{p}\left [ \erho,\prho \right ]+V_\mathrm{ee}\left [ \erho,\prho \right ] + V_\mathrm{ep}\left [ \erho,\prho \right ]+V_\mathrm{pp}\left [ \erho,\prho \right ].
\end{align}
Wherein, $\Psi_{\mathrm{ep}}$ is the exact ground eigenstate of the TC Hamiltonian, and, $\erho (\br_\mathrm{e}) = N_\mathrm{e} \sum_{\mathrm{spin}} \int d \br_{2,\mathrm{e}} \cdots \int d \br_{N_\mathrm{e},\mathrm{e}} \int d \br_{1,\mathrm{p}} \cdots \int d \br_{N_\mathrm{p},\mathrm{p}} | \Psi_{\mathrm{ep}} |^2 $ and 
$\prho (\br_\mathrm{p}) = N_\mathrm{p} \sum_{\mathrm{spin}} \int d \br_{1,\mathrm{e}} \cdots \int d \br_{N_\mathrm{e},\mathrm{e}} \int d \br_{2,\mathrm{p}} \cdots \int d \br_{N_\mathrm{p},\mathrm{p}} | \Psi_{\mathrm{ep}} |^2 $ are the one-electron and one-PCP densities, respectively. Note that $F \left [\erho,\prho \right ]$ is a universal functional of the two one-particle densities independent from the nature of the external potentials. Assuming a TC non-interacting KS reference system, the universal functional may be rewritten accordingly (for details see Ref. \cite{chakraborty_properties_2009}):
\begin{align}\label{eq:3}
    F \left [\erho,\prho \right ] = &T_\mathrm{e}^{s}\left [ \erho \right ]+T_\mathrm{p}^{s}\left [ \prho \right ]+J_\mathrm{ee}\left [ \erho \right ]+J_\mathrm{ep}\left [ \erho,\prho \right ] + J_\mathrm{pp}\left [ \prho \right ] + E_\mathrm{xc}^{\mathrm{total}}\left [ \erho,\prho \right ].
\end{align}
In this partitioning $T_\upalpha^{s} \left [ \arho \right ] = -\frac{1}{2 m_\upalpha}\sum_{i=1}^{N_\upalpha} \langle \phi_{\upalpha,i} | \nabla_{\upalpha,i}^2 | \phi_{\upalpha,i} \rangle$ is the non-interacting KS kinetic energy of the $\upalpha$ class of particles written within the basis of the KS spin-orbitals, $\left \{\phi_{\upalpha} \right\}$. Also, $J_\mathrm{\upalpha \upalpha}\left [ \arho \right ] = \frac{q_{\upalpha}^{2}}{2} \int d \br_{\upalpha,1} \int d \br_{\upalpha,2} \frac{\arho(\br_{\upalpha,1}) \arho(\br_{\upalpha,2})}{|\br_{\upalpha,1}-\br_{\upalpha,2}|}$ is the classical repulsive Coulomb interaction term within each class of particles and, $J_\mathrm{\upalpha \upbeta}\left [ \arho, \brho \right ] = q_{\upalpha}q_{\upbeta} \int d \br_{\upalpha,1} \int d \br_{\upbeta,1} \frac{\arho(\br_{\upalpha,1}) \brho(\br_{\upbeta,1})}{|\br_{\upalpha,1}-\br_{\upbeta,1}|}$ is the classical attractive Coulomb interaction term between particles from the two classes, and what remains, $ E_\mathrm{xc}^{\mathrm{total}}\left [ \erho,\prho \right ]$, contains all types of exchange and correlation effects not included in the previous terms. Complication emerges at the next stage when one tries to partition the total exchange-correlation functional further into the three contributions mentioned previously: $ E_\mathrm{xc}^{\mathrm{total}}\left [ \erho,\prho \right ]= E_\mathrm{exc}\left [ \erho,\prho \right ]+E_\mathrm{pxc}\left [ \erho,\prho \right ]+E_\mathrm{epc}\left [ \erho,\prho \right ]$. In fact, there is no single recipe to perform the partitioning, and herein we consider two theoretically acceptable but unequal partitioning schemes (for more details see Refs. \cite{chakraborty_properties_2009,kreibich_multicomponent_2008}). By comparing Eqs. (\ref{eq:2}) and (\ref{eq:3}), one may propose the following partitioning (for the special case of a system composed of $N_\mathrm{e}$ electrons plus a single PCP see Ref. \cite{goli_twocomponent_2022}): 
\begin{align}\label{eq:4}
E_\mathrm{exc}\left [ \erho,\prho \right ] = &\left ( T_\mathrm{e}\left [ \erho,\prho \right ] - T_\mathrm{e}^{s}\left [ \erho \right ] \right )  + \left ( V_\mathrm{ee}\left [ \erho,\prho \right ] - J_\mathrm{ee}\left [ \erho \right ] \right ), \nonumber\\
E_\mathrm{pxc}\left [ \erho,\prho \right ] = &\left ( T_\mathrm{p}\left [ \erho,\prho \right ] - T_\mathrm{p}^{s}\left [ \prho \right ] \right )  + \left ( V_\mathrm{pp}\left [ \erho,\prho \right ] - J_\mathrm{pp}\left [ \prho \right ] \right ), \nonumber\\
E_\mathrm{epc}\left [ \erho,\prho \right ] = & \left ( V_\mathrm{ep}\left [ \erho,\prho \right ] - J_\mathrm{ep}\left [ \erho,\prho \right ] \right ).
\end{align}
An alternative partitioning, proposed originally by Gidopoulos \cite{gidopoulos_kohnsham_1998}, is based on assuming a Hartree product-like fictitious state, $\Phi_\mathrm{ep}=\Phi_\mathrm{e}\Phi_\mathrm{p}$, where each $\Phi_\upalpha$ contains only the variables of the $\upalpha$ class of particles and could be a fully correlated wavefunction; the best $\Phi_\mathrm{ep}$ is derivable from the energy variational principle \cite{cassam-chenai_quantum_2017}. It is straightforward to demonstrate that it leads to the following partitioning (see Ref. \cite{gidopoulos_kohnsham_1998} for details of derivation):              
\begin{align}\label{eq:5}
E_\mathrm{exc}^{\prime} \left [ \erho \right ] =  &F_\mathrm{e}\left [ \erho \right ]  - T_\mathrm{e}^{s}\left [ \erho \right ]  - J_\mathrm{ee}\left [ \erho \right ], \nonumber \\
E_\mathrm{pxc}^{\prime} \left [ \prho \right ] =  &F_\mathrm{p}\left [ \prho \right ]  - T_\mathrm{p}^{s}\left [ \prho \right ]  - J_\mathrm{pp}\left [ \prho \right ], \nonumber \\
E_\mathrm{epc}^{\prime} \left [ \erho,\prho \right ] =  &F\left [ \erho,\prho \right ]  - F_\mathrm{e}\left [ \erho \right ] -F_\mathrm{p}\left [ \prho \right ]  - J_\mathrm{ep}\left [ \erho,\prho \right ].
\end{align}
Wherein, $F_\upalpha\left [ \arho \right ] = \underset{ \Phi_{\upalpha} \rightarrow \arho}{\mathrm{Min}} \langle \Phi_{\upalpha} | \hat{T}_\upalpha+\hat{V}_\mathrm{\upalpha\upalpha} | \Phi_{\upalpha} \rangle$. In this partitioning scheme, the electron-PCP correlation functional explicitly depends on the chosen fictitious state and this is best revealed upon rewriting the functional as follows \cite{gidopoulos_kohnsham_1998}:    
\begin{align}
E_\mathrm{epc}^{\prime} \left [\erho,\prho \right ] =   \underset{ \Psi_{\mathrm{ep}} \rightarrow \erho,\prho}{\mathrm{Min}} \left \langle \Psi_{\mathrm{ep}} \left | \hat{T}_\mathrm{e}+\hat{T}_\mathrm{p}+\hat{V}_\mathrm{ee}+\hat{V}_\mathrm{ep}+\hat{V}_\mathrm{pp} \right | \Psi_{\mathrm{ep}} \right \rangle  - \underset{\substack{\Phi_{\mathrm{e}} \rightarrow \erho \\ \Phi_{\mathrm{p}} \rightarrow \prho }}{\mathrm{Min}} \left \langle \Phi_{\mathrm{ep}} \left | \hat{T}_\mathrm{e}+\hat{T}_\mathrm{p}+\hat{V}_\mathrm{ee}+\hat{V}_\mathrm{ep}+\hat{V}_\mathrm{pp} \right | \Phi_{\mathrm{ep}} \right \rangle.
\end{align}
Evidently, the two definitions of the electron-PCP correlation functionals introduced in Eqs. (\ref{eq:4}) and (\ref{eq:5}) are not equivalent. All these considerations may also be extended rather straightforwardly to the spin-polarized case although it is not considered herein. 

\subsection{Double harmonic trap model of muonium atom}
The model is described by the following TC Hamiltonian, given in atomic units:
\begin{equation}\label{eq:1}
\hat{H} = -\frac{1}{2} \nabla_{\mathrm{e}}^{2} -\frac{1}{2 m_{\upmu}} \nabla_{\upmu}^{2} - \frac{1}{ | \br_\mathrm{e} - \br_\upmu  |} + \frac{1}{2}k_{\mathrm{e}} r_{\mathrm{e}}^2 + \frac{1}{2}k_{\upmu} r_{\upmu}^2.
\end{equation}
This Hamiltonian is separable into two simpler uncoupled Hamiltonians using the following variable transformations: $\bR = (\br_\mathrm{e} + m_{\upmu} \br_\upmu) / M$ and $\br = \br_\mathrm{e} - \br_\upmu$, yielding:            
\begin{align}
\hat{H} = &\hat{H}_\bR + \hat{H}_\br, \nonumber \\
\hat{H}_\bR = & -\frac{1}{2M} \nabla_{\bR}^{2} + \frac{1}{2} M \omega^2  R^2, \nonumber \\
\hat{H}_\br = & -\frac{1}{2 \mu} \nabla_{\br}^{2} + \frac{1}{2} \mu \omega^2  r^2 - \frac{1}{r}.
\end{align}
In these Hamiltonians $M=1+m_{\upmu}$ is the total mass, $\mu=m_{\upmu}/(1+m_{\upmu})$ is the reduced mass, and $\omega=\sqrt{k_\upalpha/m_\upalpha}$ is the frequency of vibration which is equal for electron and muon. $\hat{H}_\bR$ is describing the center of mass motion within a harmonic trap and its eigenvalue problem is analytically solvable and $\hat{H}_\br$ is describing the relative motion and its eigenvalue problem is solvable numerically through applying the finite element method. Accordingly, the exact solution of the ground state of the Hamiltonian given in Eq. (\ref{eq:1}) is known. The calculations were done assuming $m_{\upmu}=207$ and $\omega=0.02$ in atomic units. Two optimal sets of seven uncontracted s-type Gaussian functions were obtained at the TC Hartree-Fock (TC-HF) level of theory for the expansion of both the electron and muon spin-orbitals. Then, seven p-type and seven d-type Cartesian Gaussian functions with the same s-type optimal exponents were added to each electronic and muonic basis set to yield the final 7s7p7d basis function set for each particle type in the TC-DFT computations in Figs. 2 and S1; the optimal exponents are offered in Table \ref{tab:s3}. In TC-HF and TC-DFT computations, the basis sets and the origin of the harmonic potential traps are placed at the center of the coordinate system.  

\clearpage
\pagebreak

\begin{table}[t]
  \centering
  \caption{Total energies $E$ ($E_h$), muonic kinetic energies $K_{\upmu}$ ($E_h$), muonic bond length expectation values $\langle r_{\upmu} \rangle$ (\r{A}), and RMSDs of one-muon densities (a.u.) computed at the B3LYP/pc-2//DYPZPY/14s14p14d level of theory.}
  \begin{ruledtabular}
    \begin{tabular}{l d{4.4}d{1.4}d{1.3}d{1.3}}
     System & \multicolumn{1}{c}{$E$} & \multicolumn{1}{c}{$K_{\upmu}$}    & \multicolumn{1}{c}{$\langle r_{\upmu} \rangle$}  & \multicolumn{1}{c}{RMSD}   \\ \hline
    1 & -77.7945 & 0.0471 & 1.133 & 0.158 \\
    2 & -111.1332 & 0.0465 & 1.049 & 0.156 \\
    3 & -79.0435 & 0.0472 & 1.151 & 0.150 \\
    4 & -114.9548 & 0.0461 & 1.163 & 0.152 \\
    5 & -114.9679 & 0.0450 & 1.001 & 0.160 \\
    6 & -170.3101 & 0.0465 & 1.165 & 0.149 \\
    7 & -170.3268 & 0.0446 & 1.006 & 0.156 \\
    8 & -93.8797 & 0.0464 & 1.149 & 0.159 \\
    9 & -93.8690 & 0.0454 & 1.065 & 0.154 \\
    10 & -95.0850 & 0.0469 & 1.153 & 0.150 \\
    11 & -95.0980 & 0.0465 & 1.049 & 0.159 \\
    12 & -40.4103 & 0.0463 & 1.139 & 0.146 \\
    13 & -77.2252 & 0.0450 & 1.110 & 0.160 \\
    14 & -78.4751 & 0.0464 & 1.133 & 0.150 \\
    15 & -79.7086 & 0.0466 & 1.142 & 0.145 \\
    16 & -232.0966 & 0.0465 & 1.133 & 0.149 \\
    17 & -93.3245 & 0.0443 & 1.116 & 0.157 \\
    18 & -115.6245 & 0.0442 & 1.002 & 0.161 \\
    19 & -95.7481 & 0.0457 & 1.057 & 0.148 \\
    20 & -94.5240 & 0.0455 & 1.069 & 0.144 \\
    \end{tabular}%
    \end{ruledtabular}
  \label{tab:s1}%
\end{table}%
\clearpage

\afterpage{\onecolumngrid
\begin{longtable}[e]{@{\extracolsep{\fill}} ll d{1.4}d{1.4}d{1.4} ll d{1.4}d{1.4}d{1.4} }
\caption{Optimized geometries (\r{A}) of the muonic benchmark systems obtained at the B3LYP/pc-2//DYPZPY/14s14p14d level of theory. The Mu coordinates indicate the positions of the muonic basis set in space.}\label{tab:s2}\\
\hline\hline
System  & Atom  & \multicolumn{1}{c}{\text{x}} & \multicolumn{1}{c}{\text{y}} & \multicolumn{1}{c}{\text{z}} & System  & Atom  & \multicolumn{1}{c}{\text{x}} & \multicolumn{1}{c}{\text{y}} & \multicolumn{1}{c}{\text{z}}  \\
\hline
\endfirsthead
\caption*{\tablename{} \ref{tab:s2}. \textit{(continued)}}\\
\hline\hline
System  & Atom  & \multicolumn{1}{c}{\text{x}} & \multicolumn{1}{c}{\text{y}} & \multicolumn{1}{c}{\text{z}} & System  & Atom  & \multicolumn{1}{c}{\text{x}} & \multicolumn{1}{c}{\text{y}} & \multicolumn{1}{c}{\text{z}}  \\
\hline
\endhead
\hline
\multicolumn{10}{r}{\textit{(Table continued)}}\\
\endfoot
\hline\hline
\endlastfoot

    1     &       &       &       &       & 12    &       &       &       &  \\
          & C     & 0.0000 & 0.0000 & 0.0000 &       & C     & 0.0000 & 0.0000 & 0.0000 \\
          & C     & -1.1004 & 0.0000 & -0.6923 &       & H     & -0.5131 & -0.8889 & -0.3621 \\
          & H     & -1.4159 & 0.0000 & -1.7233 &       & H     & -0.5130 & 0.8889 & -0.3621 \\
          & H     & 0.9824 & 0.0000 & -0.4769 &       & H     & 1.0264 & 0.0000 & -0.3621 \\
          & Mu    & 0.0000 & 0.0000 & 1.1196 &       & Mu    & 0.0000 & 0.0000 & 1.1239 \\
    2     &       &       &       &       & 13    &       &       &       &  \\
          & N     & -1.0380 & -0.6512 & -0.5640 &       & C     & 0.0000 & 0.0000 & 0.0000 \\
          & H     & -0.8101 & -0.7481 & -1.5549 &       & C     & 0.0000 & 0.0000 & -1.1978 \\
          & N     & 0.0000 & 0.0000 & 0.0000 &       & H     & 0.0000 & 0.0000 & -2.2601 \\
          & H     & 0.9113 & 0.0000 & -0.4398 &       & Mu    & 0.0000 & 0.0000 & 1.0924 \\
          & Mu    & 0.0000 & 0.0000 & 1.0285 & 14    &       &       &       &  \\
    3     &       &       &       &       &       & C     & 0.0000 & 0.0000 & 0.0000 \\
          & C     & 0.0000 & 0.0000 & 0.0000 &       & C     & 1.1236 & 0.0000 & -0.7026 \\
          & H     & -0.5639 & -0.8853 & -0.2996 &       & H     & -0.9677 & 0.0000 & -0.4865 \\
          & H     & -0.5639 & 0.8853 & -0.2996 &       & H     & 1.1189 & 0.0000 & -1.7860 \\
          & C     & 1.3725 & 0.0000 & -0.5533 &       & H     & 2.0965 & 0.0000 & -0.2270 \\
          & H     & 1.9136 & 0.9240 & -0.6988 &       & Mu    & 0.0000 & 0.0000 & 1.1185 \\
          & H     & 1.9136 & -0.9240 & -0.6989 & 15    &       &       &       &  \\
          & Mu    & 0.0000 & 0.0000 & 1.1374 &       & C     & 1.4193 & 0.0000 & -0.5641 \\
    4     &       &       &       &       &       & H     & 1.9774 & -0.8798 & -0.2398 \\
          & O     & 1.3065 & 0.0000 & -0.3539 &       & H     & 1.4137 & 0.0000 & -1.6557 \\
          & C     & 0.0000 & 0.0000 & 0.0000 &       & H     & 1.9774 & 0.8798 & -0.2398 \\
          & H     & -0.5411 & 0.9101 & -0.2931 &       & C     & 0.0000 & 0.0000 & 0.0000 \\
          & H     & -0.5411 & -0.9101 & -0.2931 &       & H     & -0.5573 & -0.8798 & -0.3253 \\
          & Mu    & 0.0000 & 0.0000 & 1.1487 &       & H     & -0.5574 & 0.8798 & -0.3253 \\
    5     &       &       &       &       &       & Mu    & 0.0000 & 0.0000 & 1.1278 \\
          & C     & 1.2746 & 0.0000 & -0.4842 & 16    &       &       &       &  \\
          & H     & 2.0837 & 0.3384 & 0.1487 &       & C     & 0.0026 & 1.2041 & -2.0900 \\
          & H     & 1.3365 & 0.0623 & -1.5587 &       & C     & 0.0024 & 0.0000 & -2.7859 \\
          & O     & 0.0000 & 0.0000 & 0.0000 &       & C     & 0.0006 & -1.2041 & -2.0900 \\
          & Mu    & 0.0000 & 0.0000 & 0.9723 &       & C     & -0.0002 & -1.2039 & -0.6986 \\
    6     &       &       &       &       &       & C     & 0.0000 & 0.0000 & 0.0000 \\
          & N     & 1.3593 & 0.0000 & -0.4802 &       & C     & 0.0010 & 1.2039 & -0.6986 \\
          & H     & 1.8199 & -0.8778 & -0.2746 &       & H     & 0.0044 & 2.1418 & -2.6305 \\
          & H     & 1.3940 & 0.1432 & -1.4818 &       & H     & 0.0040 & 0.0000 & -3.8681 \\
          & O     & -0.8278 & -1.0155 & -0.3336 &       & H     & 0.0004 & -2.1418 & -2.6304 \\
          & C     & 0.0000 & 0.0000 & 0.0000 &       & H     & -0.0002 & -2.1430 & -0.1607 \\
          & H     & -0.4780 & 0.9637 & -0.2434 &       & H     & 0.0007 & 2.1430 & -0.1607 \\
          & Mu    & 0.0000 & 0.0000 & 1.1505 &       & Mu    & 0.0000 & 0.0000 & 1.1190 \\
    7     &       &       &       &       & 17    &       &       &       &  \\
          & N     & 2.1950 & -0.6677 & 0.4408 &       & C     & 0.0000 & 0.0000 & 0.0000 \\
          & H     & 2.0926 & -1.6822 & 0.4662 &       & N     & 0.0000 & 0.0000 & -1.1470 \\
          & H     & 3.1608 & -0.4313 & 0.2687 &       & Mu    & 0.0000 & 0.0000 & 1.0966 \\
          & C     & 1.3009 & 0.0000 & -0.4134 & 18    &       &       &       &  \\
          & H     & 1.3847 & -0.0280 & -1.4961 &       & C     & 1.3370 & 0.0000 & -0.4816 \\
          & O     & 0.0000 & 0.0000 & 0.0000 &       & H     & 1.2833 & 0.0006 & -1.5692 \\
          & Mu    & 0.0000 & 0.0000 & 0.9776 &       & H     & 1.8921 & 0.8896 & -0.1660 \\
    8     &       &       &       &       &       & H     & 1.8915 & -0.8901 & -0.1666 \\
          & C     & 0.0000 & 0.0000 & 0.0000 &       & O     & 0.0000 & 0.0000 & 0.0000 \\
          & H     & -0.9800 & 0.0000 & -0.4954 &       & Mu    & 0.0000 & 0.0000 & 0.9722 \\
          & N     & 1.0543 & 0.0000 & -0.6442 & 19    &       &       &       &  \\
          & Mu    & 0.0000 & 0.0000 & 1.1351 &       & C     & 1.3583 & 0.0000 & -0.5441 \\
    9     &       &       &       &       &       & H     & 1.3111 & 0.0264 & -1.6338 \\
          & C     & 1.0635 & 0.0000 & -0.6115 &       & H     & 1.8757 & 0.9055 & -0.2247 \\
          & H     & 1.1647 & 0.0000 & -1.7032 &       & H     & 1.9778 & -0.8595 & -0.2563 \\
          & N     & 0.0000 & 0.0000 & 0.0000 &       & N     & 0.0000 & 0.0000 & 0.0000 \\
          & Mu    & 0.0000 & 0.0000 & 1.0448 &       & H     & -0.5098 & -0.8210 & -0.2997 \\
    10    &       &       &       &       &       & Mu    & 0.0000 & 0.0000 & 1.0337 \\
          & N     & 1.3463 & 0.0000 & -0.4881 & 20    &       &       &       &  \\
          & H     & 1.7384 & -0.9304 & -0.3152 &       & C     & 1.1709 & 0.0000 & -0.4735 \\
          & C     & 0.0000 & 0.0000 & 0.0000 &       & H     & 1.3019 & 0.0000 & -1.5565 \\
          & H     & -0.5277 & 0.8970 & -0.3231 &       & H     & 2.0921 & 0.0000 & 0.1182 \\
          & H     & -0.5752 & -0.8888 & -0.2903 &       & N     & 0.0000 & 0.0000 & 0.0000 \\
          & Mu    & 0.0000 & 0.0000 & 1.1394 &       & Mu    & 0.0000 & 0.0000 & 1.0477 \\
    11    &       &       &       &       &       &       &       &       &  \\
          & C     & 1.2410 & 0.0000 & -0.6282 &       &       &       &       &  \\
          & H     & 2.0196 & 0.5969 & -0.1781 &       &       &       &       &  \\
          & H     & 1.2503 & -0.1485 & -1.6975 &       &       &       &       &  \\
          & N     & 0.0000 & 0.0000 & 0.0000 &       &       &       &       &  \\
          & H     & -0.6834 & -0.6362 & -0.3794 &       &       &       &       &  \\
          & Mu    & 0.0000 & 0.0000 & 1.0284 &       &       &       &       &  \\ 
\end{longtable}
}

\clearpage
\pagebreak

\begin{table}[htbp]
  \centering
  \caption{Optimized exponents of the s-type Gaussian functions for Mu in the double harmonic trap model obtained at the TC-HF level of theory.}
  \begin{ruledtabular}
    \begin{tabular}{c d{2.4}d{2.4}}
    \multicolumn{1}{c}{Exponent} & \multicolumn{1}{c}{Electron} & \multicolumn{1}{c}{Muon} \\ \hline
    1     & 7.2212 & 15.8560 \\
    2     & 3.2589 & 10.4040 \\
    3     & 0.6472 & 7.0649 \\
    4     & 0.0574 & 5.2960 \\
    5     & 1.4523 & 4.1400 \\
    6     & 0.1287 & 2.0700 \\
    7     & 0.2879 & 1.0350 \\
    \end{tabular}%
    \end{ruledtabular}
  \label{tab:s3}%
\end{table}%
\clearpage

\begin{figure*}[ht]
\includegraphics[width=\textwidth]{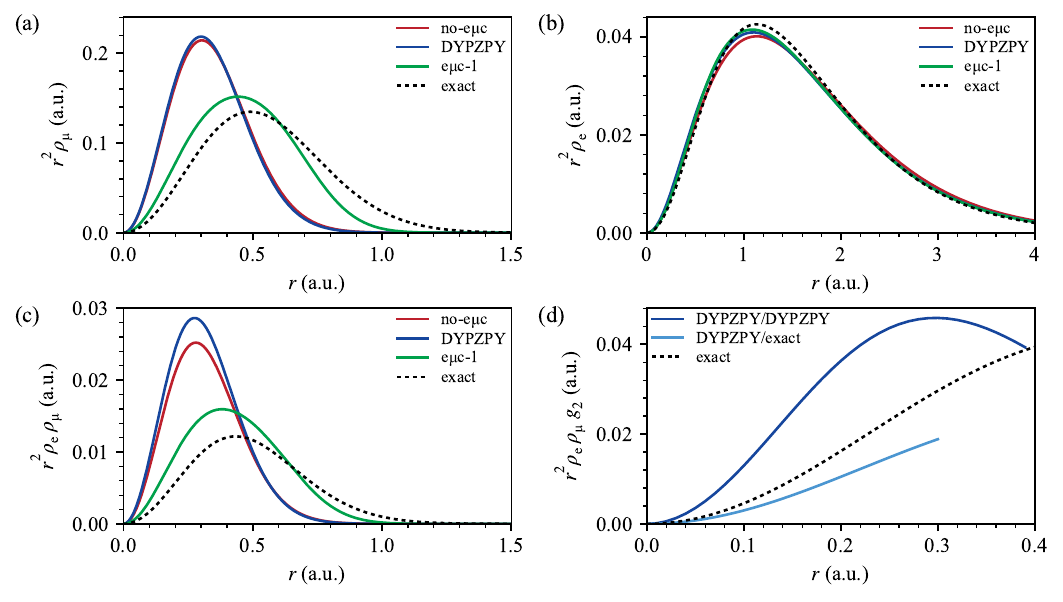}%
\caption{Radial distribution of the exact and the TC-DFT computed results for the muonium atom in the double-harmonic trap model. (a) $r^2 \mrho$, (b) $r^2 \erho$, (c) the radial distribution of the on-top density product, $r^2 \erho \mrho$, (d) the radial distribution of the on-top product $r^2 \erho \mrho g_2$ computed from the exact $\mrho$, indicated by DYPZPY/exact, and from $\mrho$ derived employing the DYPZPY functional, indicated by DYPZPY/ DYPZPY.}
\label{fig:s1}
\end{figure*}

\clearpage

%
%